\documentclass[12pt,preprint]{aastex}
\usepackage{txfonts}
\usepackage{amssymb}
\usepackage{lscape}

\shorttitle{An IC/CMB Model for the 3C 273 X-ray Jet} \shortauthors{Liu W.-P.}

\begin{document}


\title{AN IC/CMB INTERPRETATION FOR THE LARGE-SCALE JET X-RAY EMISSION OF 3C 273}


\author{Wen-Po ~Liu\altaffilmark{1,2}}


\altaffiltext{1}{Correspondence author; College of Science, Civil Aviation University of China, Tianjin 300300, China; email: wp-liu@cauc.edu.cn}
\altaffiltext{2}{Key Laboratory for Research in Galaxies and Cosmology, Chinese Academy of Sciences, 96 Jinzhai Road, Hefei 230026, Anhui, China}


\begin{abstract}

We present that the model of inverse Compton scattering of cosmic microwave background photons (IC/CMB) could well explain the large-scale jet X-ray radiation of 3C 273, and does not violate new $\emph{Fermi}$ observations. For the individual knots, the synchrotron spectrum of the low-energy electrons responsible for the IC/CMB X-ray emission may be different from the extrapolation of the 10 GHz radio spectrum of knots. Based on the IC/CMB model for the 3C 273 large-scale jet, the $\emph{Fermi}$ observations may mainly come from the small-scale jet of 3C 273 (i.e., the core). Future observations could examine our interpretation on the spectral energy distributions (SED) of knots and large-scale jet in 3C 273.

\end{abstract}



\section{INTRODUCTION}
The jet of 3C 273 is nearby (z = 0.158) large-scale jet (a possible deprojected length is $\sim$ 0.654 Mpc, Harris \& Krawczynski 2006), this ideal lab of the large-scale jet has been studied through multi-band observations (e.g., Jester et al. 2001, 2005, 2006, 2007; Uchiyama et al. 2006; Meyer \& Georganopoulos, 2014). The SED of the 3C 273 jet imply  a two-component nature (Sambruna et al. 2001; Jester et al. 2006; Uchiyama et al. 2006): a synchrotron low-energy component from radio to optical and a high-energy component including X-rays which of mechanism is still puzzling. Many researchers discuss the merits and demerits of possible mechanisms responsible for the X-ray jet (e.g., Sambruna et al. 2001; Uchiyama et al. 2006; see Harris \& Krawczynski 2006, for a review). Here, we only discuss IC/CMB model for the X-ray jet emission of 3C 273. The Achilles' heel of IC/CMB model for the X-ray jet emission is that this model leads to high, even super-Eddington kinetic power (Dermer \& Atoyan 2004; Uchiyama et al. 2006; Meyer \& Georganopoulos, 2014). We think this problem may be due to the symmetrical assumption of an isotropic pitch-angle distribution, an anisotropic pichi-angle distribution may be more close to real situation. For simplicity we consider only the symmetric assumpiton. Recently, Meyer \& Georganopoulos (2014) presents $\emph{Fermi}$ observations rule out the IC/CMB X-ray model entirely for knot A in the 3C 273 jet. But we have a different answer.

In $\S$ 2, we present that IC/CMB model could well explain the large-scale jet X-ray radiation of 3C 273, and apply IC/CMB model to the individual knots in the 3C 273 jet, which does not violate new $\emph{Fermi}$ observations. A conclusion is given in $\S$ 3.

\section{THE MODEL, FITTING RESULTS AND DISCUSSION}
We consider a standard synchrotron scenario (Kardashev 1962) in which source particles have a power law distribution, but we think particle spectral index may be a piecewise function, which is due to the complex and nonlinear acceleration (e.g., Liu \& Shen, 2007; Sahayanathan, 2008). For isotropic distributions, the synchrotron flux expression is (Kardashev 1962):

\begin{equation}
I_{\nu,s}\propto\left\{
\begin{array}{ll}
\nu^{-(p-1)/2},&\mbox{~$\nu_{min,s} \ll \nu \ll \nu_{s}$;}\\
\nu^{-p/2},&\mbox{~$\nu_{s} \ll \nu \ll \nu_{max,s}$,}\\
\end{array}
\right. \end{equation}

where $p$ is particle spectral index, $\nu_{max,s}$ and $\nu_{min,s}$ are the maximum and minimum cutoff frequencies of synchrotron spectrum. $\nu_{s}$ is the synchrotron break frequency, i.e., the peak frequency in synchrotron SED.

IC/CMB spectrum is an exact copy of the synchrotron one (Tavecchio et al. 2000; Celotti et al. 2001; Georganopoulos et al. 2006):
\begin{equation}
I_{\nu,c}\propto\left\{
\begin{array}{ll}
\nu^{-(p-1)/2},&\mbox{~$\nu_{min,c} \ll \nu \ll \nu_{c}$;}\\
\nu^{-p/2},&\mbox{~$\nu_{c} \ll \nu \ll \nu_{max,c}$,}\\
\end{array}
\right. \end{equation}
where $\nu_{max,c}$ and $\nu_{min,c}$ are the maximum and minimum cutoff frequencies of IC/CMB spectrum. $\nu_{c}$ is the IC/CMB break frequency, i.e., the peak frequency in IC/CMB SED. For the 3C 273 jet with equipartition conditions $B\delta\approx 10^{-4}$ G (Jester et al. 2005), $\nu_c=6.6 \times 10^8 \delta^2 \nu_s$ and $L_c=2.5 \times 10^{-3} \delta^4 L_s$ (Georganopoulos et al. 2006; Meyer \& Georganopoulos 2014), where $L_c$ and $L_s$ are the peak IC/CMB and synchrotron luminosities.

The data (from radio to X-ray) of knots and large-scale jet are compiled from Jester et al. (2005, 2007) and Uchiyama et al. (2006) (see Liu \& Shen (2009), for detail). For simplicity, we chose $\nu_{max,s} \sim \nu_{min,c} \sim \nu_{cut}$ and assume $\nu_{cut}$ is between two data points. Thus, flux formula of the 3C 273 large-scale jet from radio to X-ray has a three-section form. Liu \& Shen (2009) performed independent fitting to the two components, here we simultaneously fit to the two components. We use the fitting method of Liu \& Shen (2007, 2009) and Liu et al. (2013) to obtain the best fit. We arbitrarily divide the data points from radio to X-ray into three groups, the first two groups is synchrotron component, the third group belongs to IC/CMB component. Then, we calculate the corresponding $\chi^2_{\nu}$ for all the possible combinations, and choose a combination with a minimal $\chi^2_{\nu}$ among them as the best fit.

The best fitting results are shown in Table 1. The data points and model fits are plotted in Fig. 1, including the $\emph{Fermi}$ measurements and $95\%$, $99\%$, $99.9\%$ upper limits (Meyer $\&$ Georganopoulos 2014) and the TeV flux upper limits from shallow HESS observations (Aharonian et al. 2005; Georganopoulos et al. 2006). Our fitting could well interpret the emissions of the 3C 273 jet, and does not violate the $99\%$ and $99.9\%$ 3-10 GeV band $\emph{Fermi}$ upper limit and other $\gamma$-ray observations, i.e., $\emph{Fermi}$ observations do not rule out IC/CMB X-ray model. $\nu_{cut}$ is within $1.00\times10^{15}$ to $1.86\times10^{15}$ Hz. In the case of equipartition conditions, we can derive $\nu_{c}\sim2.82\times10^{23}$ Hz, $\delta\sim5.09$ and B $\sim19.64$ $\mu$G. Based on the IC/CMB model fits, the $\emph{Fermi}$ observations may be mainly contributed by the small-scale jet of 3C 273 (i.e., the core).

If we assume that Doppler factor, magnetic field and $\nu_{cut}$ are similar along the jet, we can impose the synchrotron radio spectrum lower than 10 GHz radio frequency and the IC/CMB $\gamma$-ray spectrum of knots in the 3C 273 jet. The results are shown in Fig. 2 (please note we don't consider possible absorption effects). For the individual knots, the synchrotron spectrum of low-energy electrons responsible for the IC/CMB X-ray emission may be different from the extrapolation of the 10 GHz radio spectrum (i.e., $p$ may be a piecewise function), which may be due to more complex acceleration, even multi-populations of electrons. The IC/CMB model for knots in the 3C 273 jet still do not violate new $\emph{Fermi}$ observations. If the low-energy radio spectrum of 3C 273 large-scale jet was similar to the one of knot A in Fig. 2, then the $95\%$ 3-10 GeV band $\emph{Fermi}$ upper limit of 3C 273 even was higher than the flux expected from IC/CMB model with a lower doppler factor. Future high-resolution observations could examine our interpretation on the SED of knots and large-scale jet in 3C 273.

\section{CONCLUSION}
We fit IC/CMB model to the large-scale jet and knots X-ray radiation of 3C 273, for the individual knots $p$ may be not a constant, and model fits do not violate new $\emph{Fermi}$ observations. Based on our model fit, the $\emph{Fermi}$ observations may mainly come from the small-scale jet of 3C 273 (i.e., the core). Our model fits could examined by future observations.

\section{ACKNOWLEDGMENT}
We acknowledge the support from the National Natural Science Foundation of China (NSFC) through grants U1231106 and the Open Research Program (14010203) of Key Laboratory for Research in Galaxies and Cosmology, Chinese Academy of Sciences. We acknowledge the Scientific Research Foundation of the Civil Aviation University of China (09QD15X).

\clearpage

  \begin{table}
\caption{Parameters of Model Fits to Radio through X-ray Data. Col.
(1): Synchrotron model for low-energy component and IC/CMB model for high-energy component. Col. (2): Particle spectral index. Col. (3): Reduced chi square. Col. (4): Synchrotron peak frequency for low-energy component. Col. (5): IC/CMB peak frequency for high-energy component which is a derived parameter. Col. (6): Derived doppler factor assuming equipartition. Col. (7): Derived magnetic field assuming equipartition.}
\begin{center} \scalebox{0.85}{\begin{tabular}{crrrrrrrrrr} \hline
\hline Model & $p$ & $\chi^2_{\nu}$ & $\nu_s$ (Hz) & $\nu_c$ (Hz) & $\delta$ & $B$ ($\mu$G)\\
\hline
  Synchrotron + IC/CMB & $2.80$ & $4.65$ & $1.65\times10^{13}$ & $2.82\times10^{23}$ & $5.09$ & $19.64$\\
\hline
\end{tabular}}\end{center}
\end{table}

\begin{figure}
\includegraphics[scale=1.5] {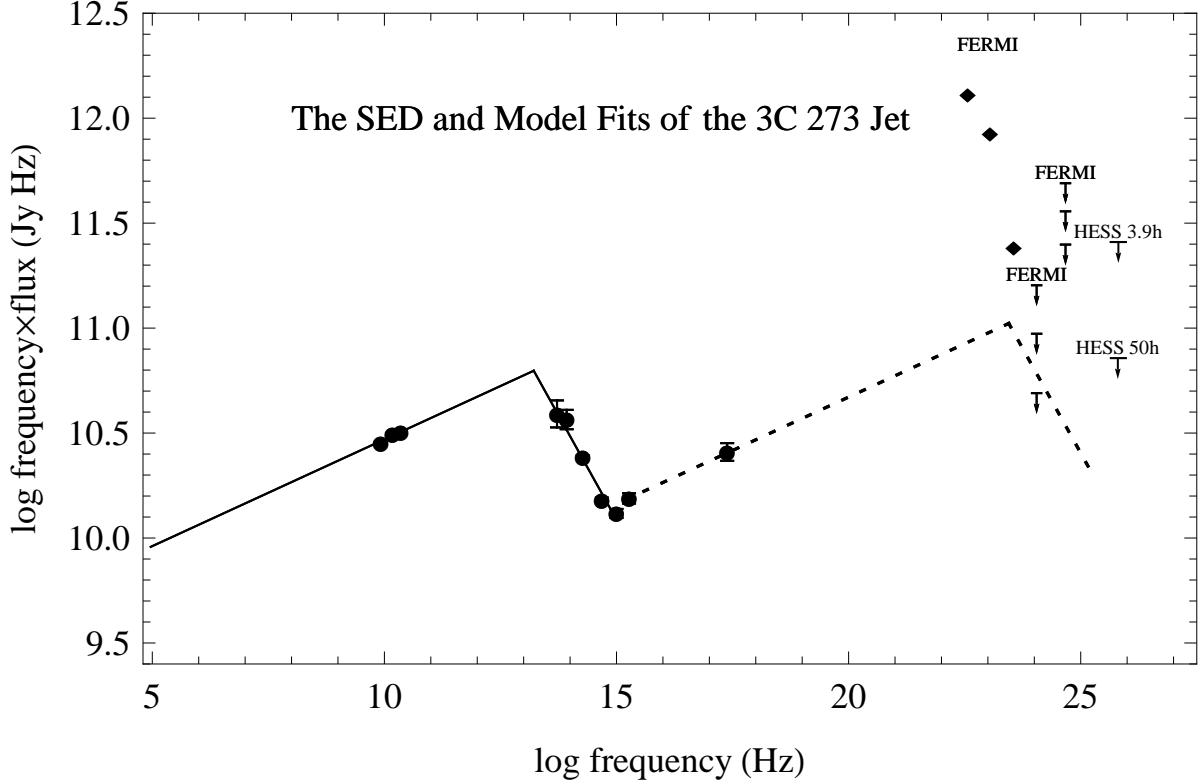}
\caption{The SED and model fits of the large-scale jet in 3C 273. The data from radio to X-ray are compiled from Jester et al. (2005, 2007) and Uchiyama et al. (2006) (see Liu \& Shen (2009), for detail), the $\emph{Fermi}$ measurements and $95\%$, $99\%$, 99.9$\%$ upper limits are from Meyer \& Georganopoulos (2014), the TeV flux upper limits from shallow HESS observations are from Aharonian et al. (2005) and Georganopoulos et al. (2006). The solid line is the parametric fit of the synchrotron low-energy component. The dashed line shows the spectrum of the IC/CMB high-energy component which does not violate the $99\%$ and $99.9\%$ 3-10 GeV band $\emph{Fermi}$ upper limit and other $\gamma$-ray observations, i.e., $\emph{Fermi}$ observations do not rule out IC/CMB X-ray model for the 3C 273 jet.}
\end{figure}

\begin{figure}
\includegraphics[scale=0.7] {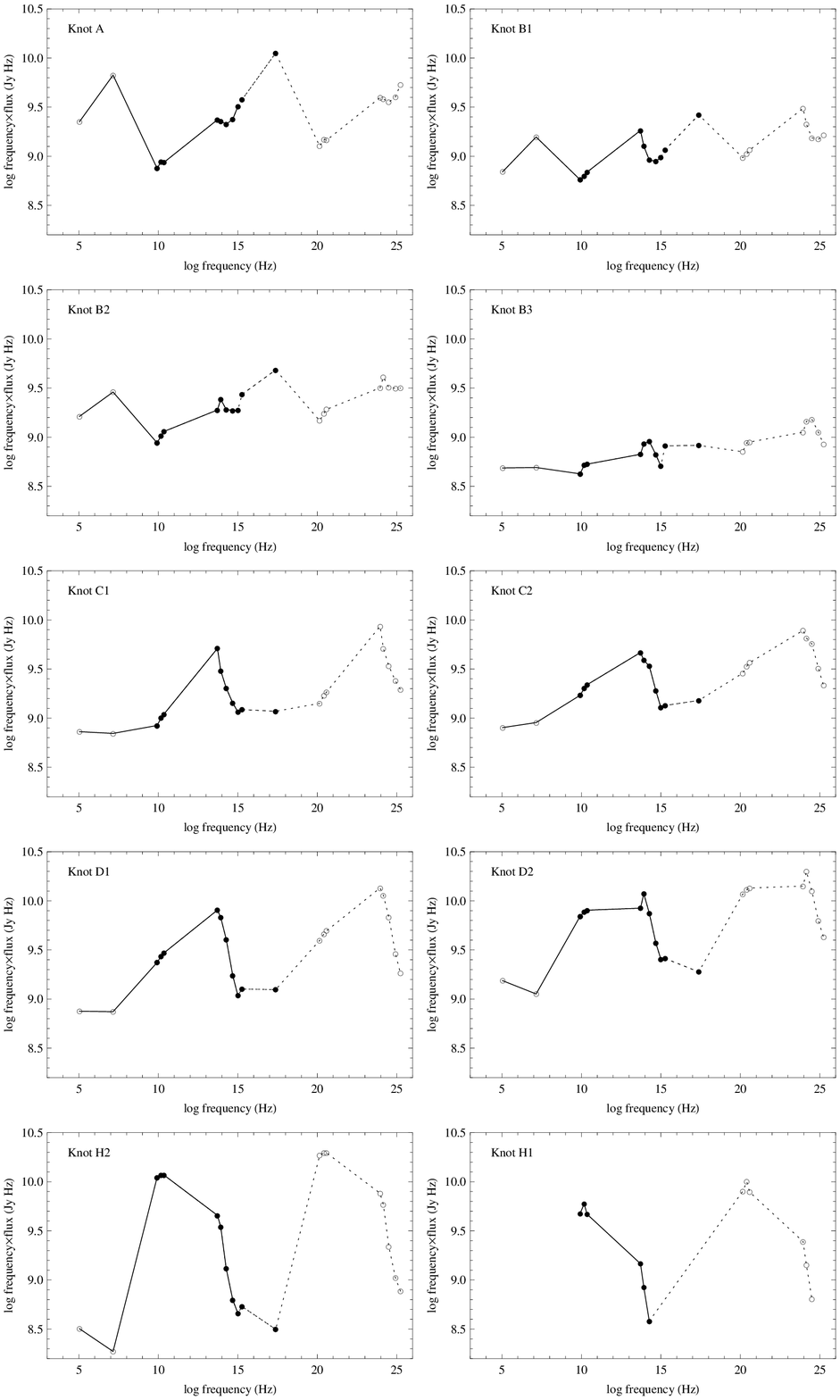}
\caption{The SED of knots along the 3C 273 jet. The observational and derived data points are plotted as filled and empty circles. The lines directly join the data points. The solid lines show synchrotron low-energy component, the dashed lines are the SED of IC/CMB high-energy component. Note we don't consider possible absorption effects.}
\end{figure}


\clearpage

\begin{references}
 \reference{1}{Aharonian, F., Akhperjanian, A. G., Bazer-Bachi, A. R., et al. 2005, A\&A, 441, 465}
 \reference{1}{Celotti, A., Ghisellini, G., \& Chiaberge, M. 2001, MNRAS, 321, L1}
  \reference{1}{Dermer, C. D. 1995, ApJ, 446, L63}
  \reference{1}{Georganopoulos, M., Perlman, E. S., Kazanas, D., \& McEnery, J.
    2006, ApJ, 653, L5}
  \reference{1}{Harris, D. E., \& Krawczynski, H. 2006, ARA\&A, 44, 463}
  \reference{1}{Jester, S., Harris, D. E., Marshall, H. L., \& Meisenheimer, K. 2006, ApJ, 648,
    900}
  \reference{1}{Jester, S., R\"{o}ser, H.-J., Meisenheimer, K., \& Perley, R. A.
    2005, A\&A, 431, 477}
  \reference{1}{Jester, S., R\"{o}ser, H.-J., Meisenheimer, K., Perley, R. A.,
    \& Conway, R. G. 2001, A\&A, 373, 447}
  \reference{1}{Jester, S. et al. 2007, MNRAS, 380, 828}
  \reference{1}{Kardashev, N. S. 1962, Soviet Astron., 6, 317}
  \reference{1}{Liu, W.-P. \& Shen, Z.-Q. 2007, ApJ, 668, L23}
   \reference{1}{Liu, W.-P. \& Shen, Z.-Q. 2009, RAA, 9, 520}
   \reference{1}{Liu, W.-P., Zhao, G.-Y., Chen, Y. J., Wang, C.-C. \& Shen, Z.-Q. 2013, AJ, 146, 155}
  \reference{1}{Meyer, E. T. \& Georganopoulos, M. 2014, ApJ, 780, L27}
  \reference{1}{Sahayanathan, S. 2008, MNRAS, 388, L49}
  \reference{1}{Sambruna, R. M., Urry, C. M., Tavecchio, F., et al. 2001, ApJ, 549, L161}
  \reference{1}{Tavecchio, F., Maraschi, L., Sambruna, R.~M., \& Urry, C.~M. 2000, ApJ, 544, L23}
  \reference{1}{Uchiyama, Y. et al. 2006, ApJ, 648, 910}

\end{references}
\end{document}